\documentclass[reprint,amsmath,amssymb,prb]{revtex4-2}

\usepackage{graphicx}
\usepackage{dcolumn}
\usepackage{bm}
\usepackage{hyperref}

\begin{document}

\author{N.~S.~Pavlov}
\email{pavlov@iep.uran.ru}
\affiliation{Institute for Electrophysics, Russian Academy of Sciences, Ekaterinburg 620016, Russia}
\affiliation{P.N. Lebedev Physical Institute, Russian Academy of Sciences, Moscow, 119991, Russia}

\author{I.~R.~Shein}
\affiliation{Institute of Solid State Chemistry, Ekaterinburg, 620108, Russia}

\author{I.~A.~Nekrasov}
\affiliation{Institute for Electrophysics, Russian Academy of Sciences, Ekaterinburg 620016, Russia}
\affiliation{P.N. Lebedev Physical Institute, Russian Academy of Sciences, Moscow, 119991, Russia}

\title{Pressurized phase transition cascade in BaMn$_2$P$_2$ and BaMn$_2$As$_2$}

\date{\today}

\begin{abstract}
	The structural analogue of iron-based superconductors the BaMn$_2$P$_2$ and BaMn$_2$As$_2$ compounds under hydrostatic pressure upto 140 GPa were studied within the framework of DFT+U.
	The transition from an antiferromagnetic (AFM) insulator to an antiferromagnetic metal is observed under pressure of 6.4 GPa for BaMn$_2$P$_2$ and 8.3 GPa for BaMn$_2$As$_2$. This second order phase transition to the AFM metallic state provides an appropriate normal state for possible superconductivity in these materials.
	Moreover, a further increase in pressure leads to a series of first order magnetostructural phase transitions between different antiferromagnetic phases, then to a ferromagnetic metal and finally to a nonmagnetic metal.
	In case of doping these compounds could potentially be a superconductors under pressure (above 6-8 GPa) with critical temperature growing under pressure.
\end{abstract}

\maketitle

\section{Introduction}
The interest to the family of iron-based high-temperature superconducting pnictides and chalcogenides (see reviews~\cite{Sadovskii_08,Stewart,Hoso_09})
gave rise to the search for new families of chemical and/or structural analogues of these systems (see, e.g.~\cite{Neupane2012,JTLRev}). One of such family are materials with complete substitution of Fe by other chemical elements, for example, manganese Mn. Since Mn is a magnetic ion due to the half filled 3d shell most of its compounds are magnetic. However, in case external pressure is applied the magnetism could be suppressed. Thus a possibility of superconductivity appears.

First superconductivity observation in Mn-based system was done in 2021 for MnSe compound with $T_c \sim 9$~K at 35~GPa~\cite{MnSe_hung_2021}.
For another compound BaMn$_2$As$_2$, structural analogue of the BaFe$_2$As$_2$, the transition to a metallic state with sharp decrease of resistivity below 17~K at 5.8~GPa~\cite{BaMn2As2_metal_2011} was experimentally observed. Magnetic measurements were not carried out in~\cite{BaMn2As2_metal_2011}. Therefore, the possibility of superconductivity in BaMn$_2$As$_2$ has not been studied in the detail up to now.
The BaMn$_2$As$_2$ is well studied material at ambient pressure(see, e.g.~\cite{BaMn2As2_Singh2009,BaMn2As2_Pandey2011,BaMn2As2_Antal2012,BaMn2As2_Zhang2016}), but under pressure it has not been investigated.
Also, the isostructural and isovalent BaMn$_2$P$_2$ compound under pressure has not been  yet studied theoretically or experimentally.

In this paper, the BaMn$_2$P$_2$ and BaMn$_2$As$_2$ compounds under external hydrostatic pressures were studied within the framework of DFT+U. The pressure dependence of thermodynamic, structural and magnetic properties of BaMn$_2$P$_2$ and BaMn$_2$As$_2$ were obtained from zero pressure upto 140~GPa.
The second order phase transition from an antiferromagnetic (AFM) insulator to an AFM metal state is observed under pressure of 6.4~GPa for BaMn$_2$P$_2$ and 8.3~GPa for BaMn$_2$As$_2$. With further pressure increase a series of first order magnetostructural phase transitions between several antiferromagnetic metallic phases, then into a ferromagnetic metal and finally into a nonmagnetic metal is found. The antiferromagnetic phases could potentially be a superconducting ones under pressure (above 6-8 GPa) with critical temperature increasing with pressure possibly up to the typical value corresponding to iron pnictides.

\section{Computational Details}
The calculations were performed in the DFT+U approximation within the VASP software package~\cite{vasp}. The generalized gradient approximation (GGA) in the form of the Perdew-Burke-Ernzerhof (PBE) exchange-correlation functional~\cite{DFT_PBE} was employed. The strong onsite Coulomb repulsion of Mn-$3d$ electrons was described with the DFT+U scheme with the Dudarev approach~\cite{DFT_U}. The U values were taken: $U=1.4$~eV for BaMn$_2$As$_2$ and $U=1.2$~eV for BaMn$_2$P$_2$. The applied hydrostatic pressure is simulated by reduction of the unit cell volume. The ion positions and lattice constants for certain volume are obtained during the DFT optimization. The Gibbs2 software package~\cite{gibbs2_1} is employed to operate with Birch–Murnaghan equation of state~\cite{Birch}.


\section{Results and Discussion}
Here we consider the most typical colinear magnetic structures for $I4/mmm$ space group of symmetry: the non-magnetic (NM), ferromagnetic (FM) and anti-ferromagnetic (AFM-A, AFM-C, AFM-G types) phases to understand which one is the ground state at a given pressure $P$ (e.g. to find a minima of total energy $E(P)$).

\begin{figure}[h]
	\includegraphics[width=0.95\linewidth]{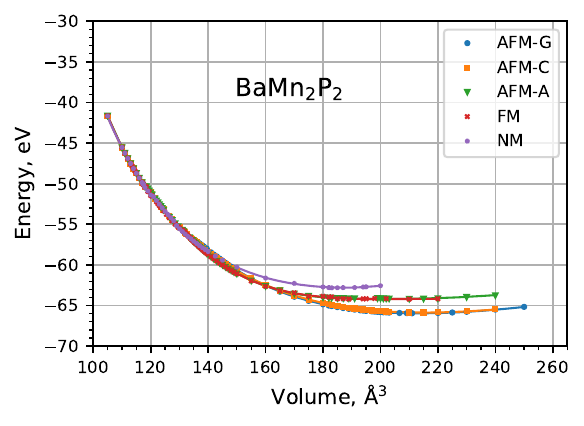}
	\includegraphics[width=0.95\linewidth]{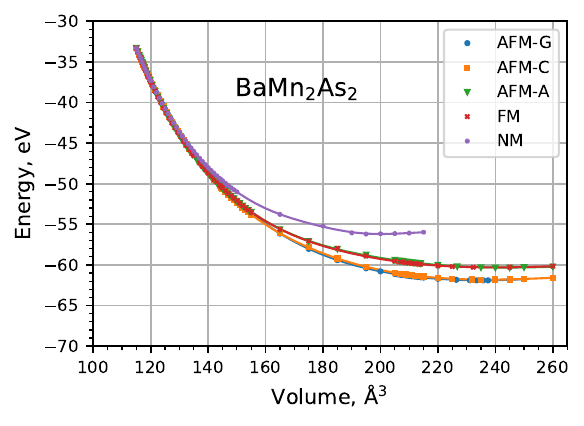}
	\caption{Total energy $E$ vs. unit cell volume $V$ for BaMn$_2$P$_2$ (top) and BaMn$_2$As$_2$ (bottom) obtained by GGA+U are plotted with symbols for all considered phases. Solid lines - fit of GGA+U data to Birch-Murnaghan equation of state.}
	\label{energy_V}
\end{figure}

From the computational point of view the unit cell volume $V$ is well defined parameter while corresponding pressure value $P$ needs to be somehow determined.
First that can be done is to calculate total energy as a function of $V$ around the minima of $E(V)$ which was estimated by full lattice DFT optimization. Corresponding results are presented in Fig.~\ref{energy_V} with symbols for all considered phases (see the legend). Solid lines are a fit of GGA+U data to Birch-Murnaghan equation of state~\cite{Birch}. Once we know the equation of state one can obtain total energy $E$ as a function of pressure $P$.

\begin{figure}[h]
	\includegraphics[width=0.95\linewidth]{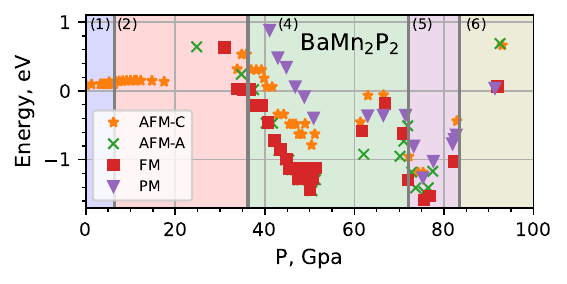}
	\includegraphics[width=0.95\linewidth]{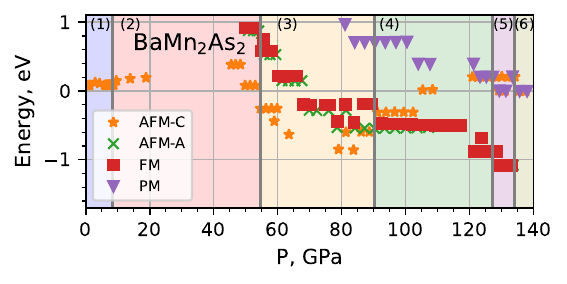}
	\caption{GGA+U total energies $E$ for BaMn$_2$P$_2$ (top) and BaMn$_2$As$_2$ (bottom) as a function of pressure $P$ with respect to AFM-G phase. (1) -- AFM-G insulator, (2) -- AFM-G metal, (3) -- AFM-C metal, (4) -- AFM-A metal, (5) -- FM metal, (6) -- non-magnetic metal. The vertical lines correspond to phase transition boundaries.}
	\label{energy_diff}
\end{figure}

In Figure~\ref{energy_diff} the $E(P)$ related to AFM-G phase (which is found to be a ground state at $P=0$~GPa) for all phases is shown. The intervals of different phases stability are separated by vertical lines. Hereafter to plot pressure dependencies of different parameters we will use $P$ values out of Birch-Murnaghan equation of state.

\begin{figure*}[ht!]
	\includegraphics[width=0.4\linewidth]{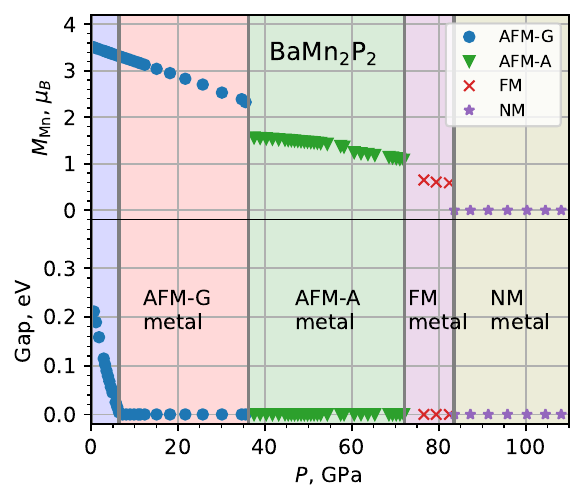}
	\includegraphics[width=0.4\linewidth]{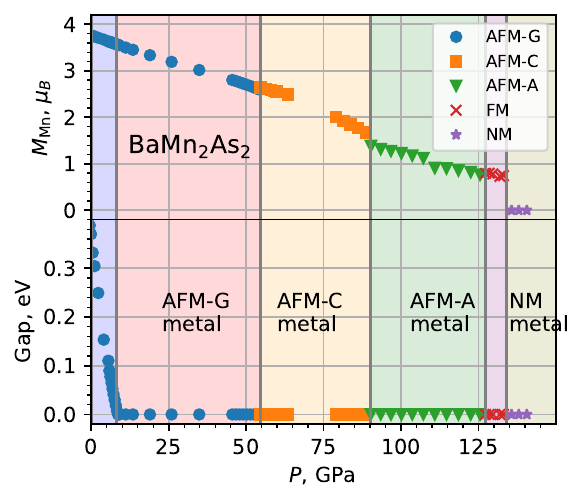}
	\caption{GGA+U dependence of Mn magnetic moment and energy gap on pressure P for BaMn$_2$P$_2$ (left) and BaMn$_2$As$_2$ (right). Here we present data only for those phases which are ground states.}
	\label{mag_gap}
\end{figure*}

The first phase transition from AFM-G insulator to AFM-G metal occurs at pressure 6.4~GPa for BaMn$_2$P$_2$ and at 8.3~GPa for BaMn$_2$As$_2$. Corresponding closing of the energy gap can be seen in Figure~\ref{mag_gap} (bottom panel). Since the gap closes continuously and magnetic order does not change there one can assume second order phase transition (Slater scenario).

Then the situation for BaMn$_2$P$_2$ and BaMn$_2$As$_2$ becomes different. At 36~GPa the AFM-A metallic solution becomes the ground state for BaMn$_2$P$_2$. Then from 72~GPa to 83~GPa BaMn$_2$P$_2$ is obtained to be a ferromagnetic metal. Above 83~GPa the Mn magnetic moment in FM phase turns to be zero (see Figure~\ref{mag_gap} top panel on the left side) and BaMn$_2$P$_2$ goes to a paramagnetic metallic ground state.

For BaMn$_2$As$_2$ the AFM-G metallic ground state undergoes to AFM-C metallic ground state at 54~GPa. Then the AFM-A phase appears to be the ground state between 90~GPa and 127~GPa. Further anti-ferromagnetism is suppressed at 127~GPa where a phase transition from metallic AFM-A to metallic FM phase occurs. Then above 134~GPa the Mn magnetic moment in FM phase disappears (see Figure~\ref{mag_gap} top panel on the right side) and the BaMn$_2$As$_2$ compound turns to a paramagnetic metal.

Let us note that there are no AFM-G solution for BaMn$_2$P$_2$ compound above 43 GPa, whereas for BaMn$_2$As$_2$ compound the AFM-G solution exists upto the transition to the nonmagnetic ground state. Also one can clearly see jumps of the Mn magnetic moment at the phase transitions boundary (see Figure~\ref{mag_gap} top panels). The values of those jumps are bigger for the phosphorous system than for the arsenic one. The change of magnetic order and presence of those jumps of magnetic moment at phase boundary let us assume first order magnetic phase transitions.

\begin{figure*}[h]
	\includegraphics[width=0.4\linewidth]{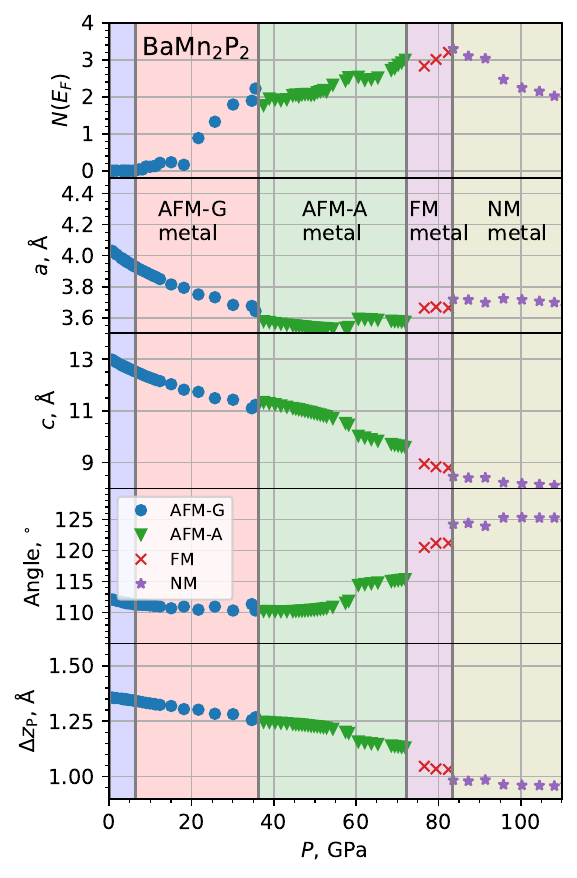}
	\includegraphics[width=0.4\linewidth]{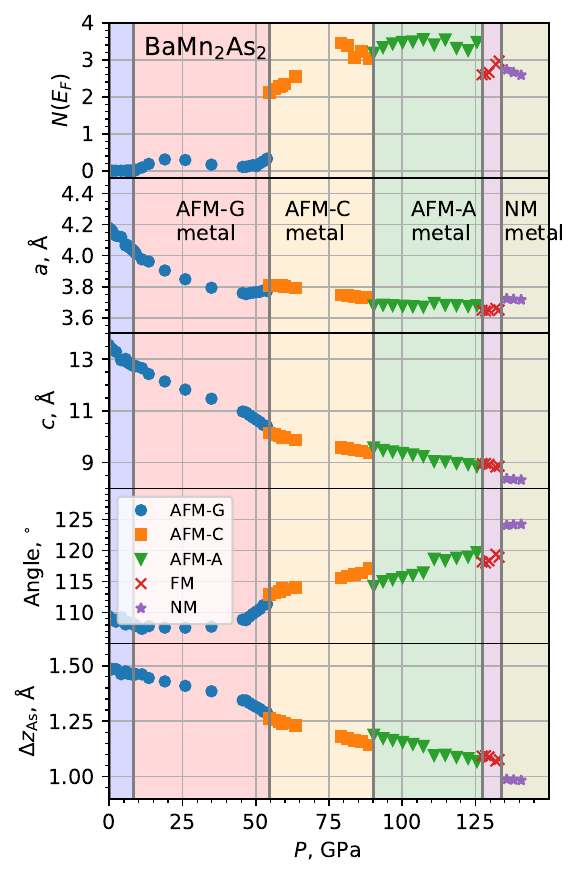}
	\caption{GGA+U dependence of total density of states at the Fermi level $N(E_F)$, lattice constants $a$ and $c$, As-Fe-As bond angle and anion height with respect to Mn plane $\Delta z$ on pressure for BaMn$_2$P$_2$ (left) and BaMn$_2$As$_2$ (right). Here we present data only for those phases which are the ground states.}
	\label{lattice_p}
\end{figure*}

To more thoroughly investigate the nature of the phase transitions we also analyze structural parameters of the materials under pressure.
In Figure~\ref{lattice_p} the GGA+U dependencies of structural parameters: $a$, $c$, As-Fe-As bond angle and anion height with respect to Mn plane $\Delta z$ on pressure for BaMn$_2$P$_2$ and BaMn$_2$As$_2$ are presented.
It was found that all observed magnetic phase transitions are accompanied by a jumps of the lattice parameters (see Figure~\ref{lattice_p}). That jumps correspond to a first order magnetostructural phase transitions. 

For BaMn$_2$P$_2$ the value of total density of states at the Fermi level $N(E_F)$ for prospective from the superconductivity existence point of view metallic AFM-G phase lays in range 0.2--2 states/eV/cell and for metallic AFM-A phase -- from 2 to 3 states/eV/cell (Figure~\ref{lattice_p}). As it is well known for iron-based superconductors $N(E_F)$ has the value from 2 to 5 states/eV/cell for paramagnetic case \cite{Kuchinskii2010_eng}. In case of BaMn$_2$As$_2$ the $N(E_F)$ value for metallic AFM-G phase is almost zero but for metallic AFM-C and AFM-A phases is large enough: 2-3 states/eV/cell. One can clearly see the uptrend of $N(E_F)$ in pressure for both materials BaMn$_2$P$_2$ and BaMn$_2$As$_2$. Therefore, in principle, one can expect that under pressure (unfortunately, quite large), the superconducting $T_C$ for AFM manganese pnictides can be of the order of magnitude, but less than for iron-based pnictides.

The  pressure dependence $N(E_F)$ on As-Fe-As bond angle $\angle$As-Fe-As and anion height with respect to Mn plane $\Delta z$ (Figure~\ref{lattice_p}) is quite different from those for iron-based pnictides without pressure \cite{Kuchinskii2010_eng}. Under pressure the $N(E_F)$ has maxima for $\angle$As-Fe-As and $\Delta z$ far away from ideal ones 109.5$^{\circ}$ and 1.37~\AA, at which the maximum of $T_c$ is reached for iron pnictides \cite{Kuchinskii2010_eng}.

However overall pressure dependence of $\angle$As-Fe-As and $\Delta z$ qualitatively agrees rather well for those of iron-based materials under pressure (see $e.g.$~\cite{KobayashiPressure2016}). Also for iron pnictides is quite typical a non-motonic behavior of $T_c$ with respect to pressure: there is some growth of $T_c$ upto some certain pressure and then $T_c$ goes down (see $e.g.$~\cite{Sefat2011}). The same behavior might be expected for manganese materials under consideration because of similar non-monotonic behavior of their $N(E_F)$ (see Figure~\ref{lattice_p}).

\section{Conclusion}
In this paper, the BaMn$_2$P$_2$ and BaMn$_2$As$_2$ compounds under external pressures were studied within the framework of DFT+U. Thermodynamic, structural and magnetic properties of BaMn$_2$P$_2$ and BaMn$_2$As$_2$ are presented at pressure from zero to 140~GPa.

The second order phase transition from an AFM-G insulator to a AFM-G metal ground state is observed under pressure of 6.4~GPa for BaMn$_2$P$_2$ and 8.3~GPa for BaMn$_2$As$_2$. Our computational results for BaMn$_2$As$_2$ suggests that experimentally observed in Ref.~\cite{BaMn2As2_metal_2011} metallic state below 17~K at 5.8~GPa could be an antiferromagnetic one.

Then the cascade of first order magnitostructural phase transitions: AFM-G metallic phase, AFM-A metallic phase, FM metallic phase and finally nonmagnetic metallic phase for BaMn$_2$P$_2$ are occurred at 36~GPa, 72~GPa and 83~GPa, correspondingly. 

The phase transition cascade for BaMn$_2$As$_2$ under pressure is slightly different: AFM-G metal, AFM-C metal, AFM-A metal, FM metal and to nonmagnetic metal take place at 54~GPa, 90~GPa, 127~GPa and 134~GPa, respectively.

Also we obtained for manganese materials under consideration  non-monotonic pressure behavior of their $N(E_F)$: there is some growth of $T_c$ upto some certain pressure and then $N(E_F)$ goes down. Once we suppose that  $T_c$ and $N(E_F)$ are connected in a BCS manner one can expect similar non-motonic behavior of $T_c$ with respect to pressure for manganese systems as those experimentally observed for iron pnictides (see $e.g.$~\cite{Sefat2011}). Since $N(E_F)$ values are nearly the same either for considered manganese materials and typical iron pnictides, in principle, one can expect that under pressure (unfortunately, quite large), the superconducting $T_c$ for manganese pnictides can be of the order of magnitude, but less (due to its AFM normal state) than for iron-based pnictides.

\section*{Acknowledgments}
The work support of the State Assignmentes \#124022200005-2 of Institute of Electrophysics and \#124020600024-5 of Institute of Solid State Chemistry UB RAS.
We are grateful to E.Z.~Kuchinskii and P.A. Igoshev for useful discussions.

\section*{Appendix}
\begin{figure*}[h]
	\includegraphics[width=0.4\linewidth]{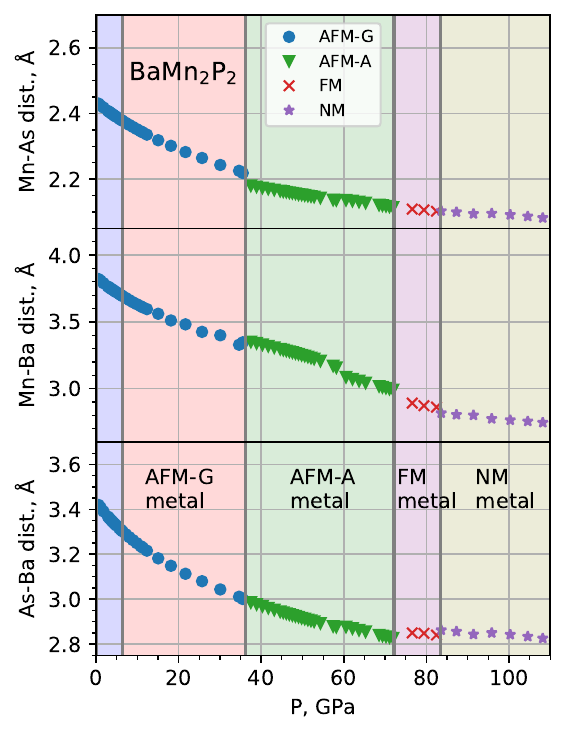}
	\includegraphics[width=0.4\linewidth]{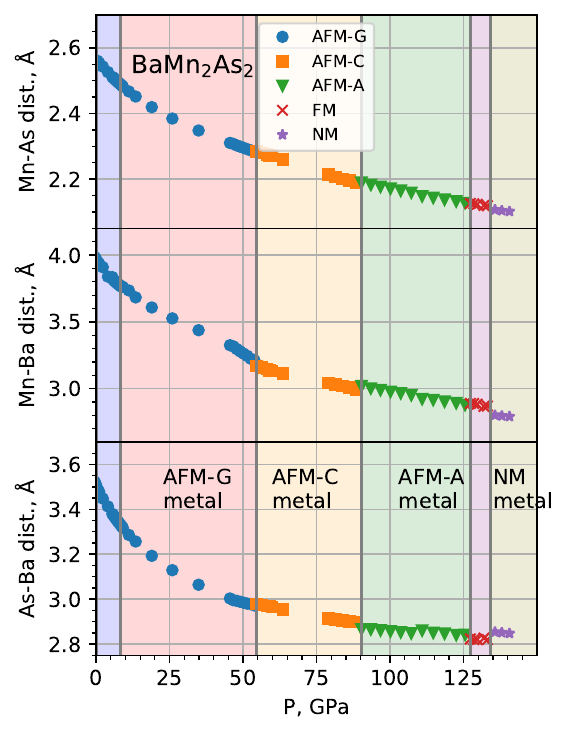}
	\caption{GGA+U dependence of different interatomit distances on pressure for BaMn$_2$P$_2$ (left) and BaMn$_2$As$_2$ (right). Here we present data only for those phases which are ground states.}
	\label{distance_p}
\end{figure*}

For completeness, the Figure~\ref{distance_p} shows pressure dependence of different interatomic distances on pressure for BaMn$_2$P$_2$ and BaMn$_2$As$_2$. Our data qualitatively agrees rather well for those of iron-based materials under pressure (see $e.g.$~\cite{KobayashiPressure2016}).
 
\bibliography{./bib_file}

\begin{thebibliography}{19}%
\makeatletter
\providecommand \@ifxundefined [1]{%
 \@ifx{#1\undefined}
}%
\providecommand \@ifnum [1]{%
 \ifnum #1\expandafter \@firstoftwo
 \else \expandafter \@secondoftwo
 \fi
}%
\providecommand \@ifx [1]{%
 \ifx #1\expandafter \@firstoftwo
 \else \expandafter \@secondoftwo
 \fi
}%
\providecommand \natexlab [1]{#1}%
\providecommand \enquote  [1]{``#1''}%
\providecommand \bibnamefont  [1]{#1}%
\providecommand \bibfnamefont [1]{#1}%
\providecommand \citenamefont [1]{#1}%
\providecommand \href@noop [0]{\@secondoftwo}%
\providecommand \href [0]{\begingroup \@sanitize@url \@href}%
\providecommand \@href[1]{\@@startlink{#1}\@@href}%
\providecommand \@@href[1]{\endgroup#1\@@endlink}%
\providecommand \@sanitize@url [0]{\catcode `\\12\catcode `\$12\catcode
  `\&12\catcode `\#12\catcode `\^12\catcode `\_12\catcode `\%12\relax}%
\providecommand \@@startlink[1]{}%
\providecommand \@@endlink[0]{}%
\providecommand \url  [0]{\begingroup\@sanitize@url \@url }%
\providecommand \@url [1]{\endgroup\@href {#1}{\urlprefix }}%
\providecommand \urlprefix  [0]{URL }%
\providecommand \Eprint [0]{\href }%
\providecommand \doibase [0]{http://dx.doi.org/}%
\providecommand \selectlanguage [0]{\@gobble}%
\providecommand \bibinfo  [0]{\@secondoftwo}%
\providecommand \bibfield  [0]{\@secondoftwo}%
\providecommand \translation [1]{[#1]}%
\providecommand \BibitemOpen [0]{}%
\providecommand \bibitemStop [0]{}%
\providecommand \bibitemNoStop [0]{.\EOS\space}%
\providecommand \EOS [0]{\spacefactor3000\relax}%
\providecommand \BibitemShut  [1]{\csname bibitem#1\endcsname}%
\let\auto@bib@innerbib\@empty
\bibitem [{\citenamefont {Sadovskii}(2008)}]{Sadovskii_08}%
  \BibitemOpen
  \bibfield  {author} {\bibinfo {author} {\bibfnamefont {M.~V.}\ \bibnamefont
  {Sadovskii}},\ }\href {https://doi.org/10.1070/PU2008v051n12ABEH006820}
  {\bibfield  {journal} {\bibinfo  {journal} {Physics-Uspekhi}\ }\textbf
  {\bibinfo {volume} {51}},\ \bibinfo {pages} {1201} (\bibinfo {year}
  {2008})},\ \bibinfo {note} {[{\em Usp. Fiz. Nauk} {\bf 2008}, {\em 178},
  1243-1271]}\BibitemShut {NoStop}%
\bibitem [{\citenamefont {Stewart}(2011)}]{Stewart}%
  \BibitemOpen
  \bibfield  {author} {\bibinfo {author} {\bibfnamefont {G.~R.}\ \bibnamefont
  {Stewart}},\ }\href {https://doi.org/10.1103/RevModPhys.83.1589} {\bibfield
  {journal} {\bibinfo  {journal} {Reviews of Modern Physics}\ }\textbf
  {\bibinfo {volume} {83}},\ \bibinfo {pages} {1589} (\bibinfo {year}
  {2011})}\BibitemShut {NoStop}%
\bibitem [{\citenamefont {Ishida}\ \emph {et~al.}(2009)\citenamefont {Ishida},
  \citenamefont {Nakai},\ and\ \citenamefont {Hosono}}]{Hoso_09}%
  \BibitemOpen
  \bibfield  {author} {\bibinfo {author} {\bibfnamefont {K.}~\bibnamefont
  {Ishida}}, \bibinfo {author} {\bibfnamefont {Y.}~\bibnamefont {Nakai}}, \
  and\ \bibinfo {author} {\bibfnamefont {H.}~\bibnamefont {Hosono}},\ }\href
  {https://doi.org/10.1143/JPSJ.78.062001} {\bibfield  {journal} {\bibinfo
  {journal} {Journal of the Physical Society of Japan}\ }\textbf {\bibinfo
  {volume} {78}},\ \bibinfo {pages} {062001} (\bibinfo {year}
  {2009})}\BibitemShut {NoStop}%
\bibitem [{\citenamefont {Neupane}\ \emph {et~al.}(2012)\citenamefont
  {Neupane}, \citenamefont {Liu}, \citenamefont {Xu}, \citenamefont {Wang},
  \citenamefont {Ni}, \citenamefont {Allred}, \citenamefont {Wray},
  \citenamefont {Alidoust}, \citenamefont {Lin}, \citenamefont {Markiewicz},
  \citenamefont {Bansil}, \citenamefont {Cava},\ and\ \citenamefont
  {Hasan}}]{Neupane2012}%
  \BibitemOpen
  \bibfield  {author} {\bibinfo {author} {\bibfnamefont {M.}~\bibnamefont
  {Neupane}}, \bibinfo {author} {\bibfnamefont {C.}~\bibnamefont {Liu}},
  \bibinfo {author} {\bibfnamefont {S.-Y.}\ \bibnamefont {Xu}}, \bibinfo
  {author} {\bibfnamefont {Y.-J.}\ \bibnamefont {Wang}}, \bibinfo {author}
  {\bibfnamefont {N.}~\bibnamefont {Ni}}, \bibinfo {author} {\bibfnamefont
  {J.~M.}\ \bibnamefont {Allred}}, \bibinfo {author} {\bibfnamefont {L.~A.}\
  \bibnamefont {Wray}}, \bibinfo {author} {\bibfnamefont {N.}~\bibnamefont
  {Alidoust}}, \bibinfo {author} {\bibfnamefont {H.}~\bibnamefont {Lin}},
  \bibinfo {author} {\bibfnamefont {R.~S.}\ \bibnamefont {Markiewicz}},
  \bibinfo {author} {\bibfnamefont {A.}~\bibnamefont {Bansil}}, \bibinfo
  {author} {\bibfnamefont {R.~J.}\ \bibnamefont {Cava}}, \ and\ \bibinfo
  {author} {\bibfnamefont {M.~Z.}\ \bibnamefont {Hasan}},\ }\href {\doibase
  10.1103/PhysRevB.85.094510} {\bibfield  {journal} {\bibinfo  {journal}
  {Physical Review B}\ }\textbf {\bibinfo {volume} {85}},\ \bibinfo {pages}
  {094510} (\bibinfo {year} {2012})}\BibitemShut {NoStop}%
\bibitem [{\citenamefont {Nekrasov}\ and\ \citenamefont
  {Sadovskii}(2014)}]{JTLRev}%
  \BibitemOpen
  \bibfield  {author} {\bibinfo {author} {\bibfnamefont {I.~A.}\ \bibnamefont
  {Nekrasov}}\ and\ \bibinfo {author} {\bibfnamefont {M.~V.}\ \bibnamefont
  {Sadovskii}},\ }\href {https://doi.org/10.1134/S0021364014100075} {\bibfield
  {journal} {\bibinfo  {journal} {JETP Letters}\ }\textbf {\bibinfo {volume}
  {99}},\ \bibinfo {pages} {598} (\bibinfo {year} {2014})},\ \bibinfo {note}
  {[{\em Pis'ma Zh. Eksp. Teor. Fiz.} {\bf 2014}, {\em 99}, 687]}\BibitemShut
  {NoStop}%
\bibitem [{\citenamefont {Hung}\ \emph {et~al.}(2021)\citenamefont {Hung},
  \citenamefont {Huang}, \citenamefont {Deng}, \citenamefont {Ou},
  \citenamefont {Chen}, \citenamefont {Wu}, \citenamefont {Huyan},
  \citenamefont {Chu}, \citenamefont {Chen},\ and\ \citenamefont
  {Lee}}]{MnSe_hung_2021}%
  \BibitemOpen
  \bibfield  {author} {\bibinfo {author} {\bibfnamefont {T.~L.}\ \bibnamefont
  {Hung}}, \bibinfo {author} {\bibfnamefont {C.~H.}\ \bibnamefont {Huang}},
  \bibinfo {author} {\bibfnamefont {L.~Z.}\ \bibnamefont {Deng}}, \bibinfo
  {author} {\bibfnamefont {M.~N.}\ \bibnamefont {Ou}}, \bibinfo {author}
  {\bibfnamefont {Y.~Y.}\ \bibnamefont {Chen}}, \bibinfo {author}
  {\bibfnamefont {M.~K.}\ \bibnamefont {Wu}}, \bibinfo {author} {\bibfnamefont
  {S.~Y.}\ \bibnamefont {Huyan}}, \bibinfo {author} {\bibfnamefont {C.~W.}\
  \bibnamefont {Chu}}, \bibinfo {author} {\bibfnamefont {P.~J.}\ \bibnamefont
  {Chen}}, \ and\ \bibinfo {author} {\bibfnamefont {T.~K.}\ \bibnamefont
  {Lee}},\ }\href {\doibase 10.1038/s41467-021-25721-1} {\bibfield  {journal}
  {\bibinfo  {journal} {Nature Communications}\ }\textbf {\bibinfo {volume}
  {12}},\ \bibinfo {pages} {5436} (\bibinfo {year} {2021})}\BibitemShut
  {NoStop}%
\bibitem [{\citenamefont {Satya}\ \emph {et~al.}(2011)\citenamefont {Satya},
  \citenamefont {Mani}, \citenamefont {Arulraj}, \citenamefont {Shekar},
  \citenamefont {Vinod}, \citenamefont {Sundar},\ and\ \citenamefont
  {Bharathi}}]{BaMn2As2_metal_2011}%
  \BibitemOpen
  \bibfield  {author} {\bibinfo {author} {\bibfnamefont {A.~T.}\ \bibnamefont
  {Satya}}, \bibinfo {author} {\bibfnamefont {A.}~\bibnamefont {Mani}},
  \bibinfo {author} {\bibfnamefont {A.}~\bibnamefont {Arulraj}}, \bibinfo
  {author} {\bibfnamefont {N.~V.~C.}\ \bibnamefont {Shekar}}, \bibinfo {author}
  {\bibfnamefont {K.}~\bibnamefont {Vinod}}, \bibinfo {author} {\bibfnamefont
  {C.~S.}\ \bibnamefont {Sundar}}, \ and\ \bibinfo {author} {\bibfnamefont
  {A.}~\bibnamefont {Bharathi}},\ }\href {\doibase 10.1103/PhysRevB.84.180515}
  {\bibfield  {journal} {\bibinfo  {journal} {Phys. Rev. B}\ }\textbf {\bibinfo
  {volume} {84}},\ \bibinfo {pages} {180515} (\bibinfo {year}
  {2011})}\BibitemShut {NoStop}%
\bibitem [{\citenamefont {Singh}\ \emph {et~al.}(2009)\citenamefont {Singh},
  \citenamefont {Ellern},\ and\ \citenamefont {Johnston}}]{BaMn2As2_Singh2009}%
  \BibitemOpen
  \bibfield  {author} {\bibinfo {author} {\bibfnamefont {Y.}~\bibnamefont
  {Singh}}, \bibinfo {author} {\bibfnamefont {A.}~\bibnamefont {Ellern}}, \
  and\ \bibinfo {author} {\bibfnamefont {D.~C.}\ \bibnamefont {Johnston}},\
  }\href {\doibase 10.1103/PhysRevB.79.094519} {\bibfield  {journal} {\bibinfo
  {journal} {Physical Review B}\ }\textbf {\bibinfo {volume} {79}},\ \bibinfo
  {pages} {2} (\bibinfo {year} {2009})}\BibitemShut {NoStop}%
\bibitem [{\citenamefont {Pandey}\ \emph {et~al.}(2011)\citenamefont {Pandey},
  \citenamefont {Anand},\ and\ \citenamefont {Johnston}}]{BaMn2As2_Pandey2011}%
  \BibitemOpen
  \bibfield  {author} {\bibinfo {author} {\bibfnamefont {A.}~\bibnamefont
  {Pandey}}, \bibinfo {author} {\bibfnamefont {V.~K.}\ \bibnamefont {Anand}}, \
  and\ \bibinfo {author} {\bibfnamefont {D.~C.}\ \bibnamefont {Johnston}},\
  }\href {\doibase 10.1103/PhysRevB.84.014405} {\bibfield  {journal} {\bibinfo
  {journal} {Physical Review B}\ }\textbf {\bibinfo {volume} {84}},\ \bibinfo
  {pages} {1} (\bibinfo {year} {2011})}\BibitemShut {NoStop}%
\bibitem [{\citenamefont {Antal}\ \emph {et~al.}(2012)\citenamefont {Antal},
  \citenamefont {Knoblauch}, \citenamefont {Singh}, \citenamefont {Gegenwart},
  \citenamefont {Wu},\ and\ \citenamefont {Dressel}}]{BaMn2As2_Antal2012}%
  \BibitemOpen
  \bibfield  {author} {\bibinfo {author} {\bibfnamefont {A.}~\bibnamefont
  {Antal}}, \bibinfo {author} {\bibfnamefont {T.}~\bibnamefont {Knoblauch}},
  \bibinfo {author} {\bibfnamefont {Y.}~\bibnamefont {Singh}}, \bibinfo
  {author} {\bibfnamefont {P.}~\bibnamefont {Gegenwart}}, \bibinfo {author}
  {\bibfnamefont {D.}~\bibnamefont {Wu}}, \ and\ \bibinfo {author}
  {\bibfnamefont {M.}~\bibnamefont {Dressel}},\ }\href {\doibase
  10.1103/PhysRevB.86.014506} {\bibfield  {journal} {\bibinfo  {journal}
  {Physical Review B}\ }\textbf {\bibinfo {volume} {86}},\ \bibinfo {pages} {1}
  (\bibinfo {year} {2012})}\BibitemShut {NoStop}%
\bibitem [{\citenamefont {Zhang}\ \emph {et~al.}(2016)\citenamefont {Zhang},
  \citenamefont {Richard}, \citenamefont {Van~Roekeghem}, \citenamefont {Nie},
  \citenamefont {Xu}, \citenamefont {Zhang}, \citenamefont {Miao},
  \citenamefont {Wu}, \citenamefont {Yin}, \citenamefont {Fu}, \citenamefont
  {Kong}, \citenamefont {Qian}, \citenamefont {Wang}, \citenamefont {Fang},
  \citenamefont {Sefat}, \citenamefont {Biermann},\ and\ \citenamefont
  {Ding}}]{BaMn2As2_Zhang2016}%
  \BibitemOpen
  \bibfield  {author} {\bibinfo {author} {\bibfnamefont {W.~L.}\ \bibnamefont
  {Zhang}}, \bibinfo {author} {\bibfnamefont {P.}~\bibnamefont {Richard}},
  \bibinfo {author} {\bibfnamefont {A.}~\bibnamefont {Van~Roekeghem}}, \bibinfo
  {author} {\bibfnamefont {S.~M.}\ \bibnamefont {Nie}}, \bibinfo {author}
  {\bibfnamefont {N.}~\bibnamefont {Xu}}, \bibinfo {author} {\bibfnamefont
  {P.}~\bibnamefont {Zhang}}, \bibinfo {author} {\bibfnamefont
  {H.}~\bibnamefont {Miao}}, \bibinfo {author} {\bibfnamefont {S.~F.}\
  \bibnamefont {Wu}}, \bibinfo {author} {\bibfnamefont {J.~X.}\ \bibnamefont
  {Yin}}, \bibinfo {author} {\bibfnamefont {B.~B.}\ \bibnamefont {Fu}},
  \bibinfo {author} {\bibfnamefont {L.~Y.}\ \bibnamefont {Kong}}, \bibinfo
  {author} {\bibfnamefont {T.}~\bibnamefont {Qian}}, \bibinfo {author}
  {\bibfnamefont {Z.~J.}\ \bibnamefont {Wang}}, \bibinfo {author}
  {\bibfnamefont {Z.}~\bibnamefont {Fang}}, \bibinfo {author} {\bibfnamefont
  {A.~S.}\ \bibnamefont {Sefat}}, \bibinfo {author} {\bibfnamefont
  {S.}~\bibnamefont {Biermann}}, \ and\ \bibinfo {author} {\bibfnamefont
  {H.}~\bibnamefont {Ding}},\ }\href {\doibase 10.1103/PhysRevB.94.155155}
  {\bibfield  {journal} {\bibinfo  {journal} {Physical Review B}\ }\textbf
  {\bibinfo {volume} {94}},\ \bibinfo {pages} {1} (\bibinfo {year}
  {2016})}\BibitemShut {NoStop}%
\bibitem [{\citenamefont {Kresse}\ and\ \citenamefont
  {Furthmüller}(1996)}]{vasp}%
  \BibitemOpen
  \bibfield  {author} {\bibinfo {author} {\bibfnamefont {G.}~\bibnamefont
  {Kresse}}\ and\ \bibinfo {author} {\bibfnamefont {J.}~\bibnamefont
  {Furthmüller}},\ }\href {\doibase 10.1103/PhysRevB.54.11169} {\bibfield
  {journal} {\bibinfo  {journal} {Physical Review B}\ }\textbf {\bibinfo
  {volume} {54}},\ \bibinfo {pages} {11169} (\bibinfo {year}
  {1996})}\BibitemShut {NoStop}%
\bibitem [{\citenamefont {Perdew}\ \emph {et~al.}(1996)\citenamefont {Perdew},
  \citenamefont {Burke},\ and\ \citenamefont {Ernzerhof}}]{DFT_PBE}%
  \BibitemOpen
  \bibfield  {author} {\bibinfo {author} {\bibfnamefont {J.~P.}\ \bibnamefont
  {Perdew}}, \bibinfo {author} {\bibfnamefont {K.}~\bibnamefont {Burke}}, \
  and\ \bibinfo {author} {\bibfnamefont {M.}~\bibnamefont {Ernzerhof}},\ }\href
  {\doibase 10.1103/PhysRevLett.77.3865} {\bibfield  {journal} {\bibinfo
  {journal} {Physical Review Letters}\ }\textbf {\bibinfo {volume} {77}},\
  \bibinfo {pages} {3865} (\bibinfo {year} {1996})}\BibitemShut {NoStop}%
\bibitem [{\citenamefont {Dudarev}\ \emph {et~al.}(1998)\citenamefont
  {Dudarev}, \citenamefont {Botton}, \citenamefont {Savrasov}, \citenamefont
  {Humphreys},\ and\ \citenamefont {Sutton}}]{DFT_U}%
  \BibitemOpen
  \bibfield  {author} {\bibinfo {author} {\bibfnamefont {S.~L.}\ \bibnamefont
  {Dudarev}}, \bibinfo {author} {\bibfnamefont {G.~A.}\ \bibnamefont {Botton}},
  \bibinfo {author} {\bibfnamefont {S.~Y.}\ \bibnamefont {Savrasov}}, \bibinfo
  {author} {\bibfnamefont {C.~J.}\ \bibnamefont {Humphreys}}, \ and\ \bibinfo
  {author} {\bibfnamefont {A.~P.}\ \bibnamefont {Sutton}},\ }\href {\doibase
  10.1103/PhysRevB.57.1505} {\bibfield  {journal} {\bibinfo  {journal}
  {Physical Review B}\ }\textbf {\bibinfo {volume} {57}},\ \bibinfo {pages}
  {1505} (\bibinfo {year} {1998})}\BibitemShut {NoStop}%
\bibitem [{\citenamefont {Otero-de-la Roza}\ and\ \citenamefont
  {Luaña}(2011)}]{gibbs2_1}%
  \BibitemOpen
  \bibfield  {author} {\bibinfo {author} {\bibfnamefont {A.}~\bibnamefont
  {Otero-de-la Roza}}\ and\ \bibinfo {author} {\bibfnamefont {V.}~\bibnamefont
  {Luaña}},\ }\href {\doibase 10.1016/j.cpc.2011.04.016} {\bibfield  {journal}
  {\bibinfo  {journal} {Computer Physics Communications}\ }\textbf {\bibinfo
  {volume} {182}},\ \bibinfo {pages} {1708} (\bibinfo {year}
  {2011})}\BibitemShut {NoStop}%
\bibitem [{\citenamefont {Birch}(1947)}]{Birch}%
  \BibitemOpen
  \bibfield  {author} {\bibinfo {author} {\bibfnamefont {F.}~\bibnamefont
  {Birch}},\ }\href {\doibase 10.1103/PhysRev.71.809} {\bibfield  {journal}
  {\bibinfo  {journal} {Phys. Rev.}\ }\textbf {\bibinfo {volume} {71}},\
  \bibinfo {pages} {809} (\bibinfo {year} {1947})}\BibitemShut {NoStop}%
\bibitem [{\citenamefont {Kuchinskii}\ \emph {et~al.}(2010)\citenamefont
  {Kuchinskii}, \citenamefont {Nekrasov},\ and\ \citenamefont
  {Sadovskii}}]{Kuchinskii2010_eng}%
  \BibitemOpen
  \bibfield  {author} {\bibinfo {author} {\bibfnamefont {E.~Z.}\ \bibnamefont
  {Kuchinskii}}, \bibinfo {author} {\bibfnamefont {I.~A.}\ \bibnamefont
  {Nekrasov}}, \ and\ \bibinfo {author} {\bibfnamefont {M.~V.}\ \bibnamefont
  {Sadovskii}},\ }\href {\doibase 10.1134/S0021364010100061} {\bibfield
  {journal} {\bibinfo  {journal} {JETP Letters}\ }\textbf {\bibinfo {volume}
  {91}},\ \bibinfo {pages} {518} (\bibinfo {year} {2010})}\BibitemShut
  {NoStop}%
\bibitem [{\citenamefont {Kobayashi}\ \emph {et~al.}(2016)\citenamefont
  {Kobayashi}, \citenamefont {Yamaura}, \citenamefont {Iimura}, \citenamefont
  {Maki}, \citenamefont {Sagayama}, \citenamefont {Kumai}, \citenamefont
  {Murakami}, \citenamefont {Takahashi}, \citenamefont {Matsuishi},\ and\
  \citenamefont {Hosono}}]{KobayashiPressure2016}%
  \BibitemOpen
  \bibfield  {author} {\bibinfo {author} {\bibfnamefont {K.}~\bibnamefont
  {Kobayashi}}, \bibinfo {author} {\bibfnamefont {J.-i.}\ \bibnamefont
  {Yamaura}}, \bibinfo {author} {\bibfnamefont {S.}~\bibnamefont {Iimura}},
  \bibinfo {author} {\bibfnamefont {S.}~\bibnamefont {Maki}}, \bibinfo {author}
  {\bibfnamefont {H.}~\bibnamefont {Sagayama}}, \bibinfo {author}
  {\bibfnamefont {R.}~\bibnamefont {Kumai}}, \bibinfo {author} {\bibfnamefont
  {Y.}~\bibnamefont {Murakami}}, \bibinfo {author} {\bibfnamefont
  {H.}~\bibnamefont {Takahashi}}, \bibinfo {author} {\bibfnamefont
  {S.}~\bibnamefont {Matsuishi}}, \ and\ \bibinfo {author} {\bibfnamefont
  {H.}~\bibnamefont {Hosono}},\ }\href {\doibase 10.1038/srep39646} {\bibfield
  {journal} {\bibinfo  {journal} {Scientific Reports}\ }\textbf {\bibinfo
  {volume} {6}},\ \bibinfo {pages} {39646} (\bibinfo {year}
  {2016})}\BibitemShut {NoStop}%
\bibitem [{\citenamefont {Sefat}(2011)}]{Sefat2011}%
  \BibitemOpen
  \bibfield  {author} {\bibinfo {author} {\bibfnamefont {A.~S.}\ \bibnamefont
  {Sefat}},\ }\href {\doibase 10.1088/0034-4885/74/12/124502} {\bibfield
  {journal} {\bibinfo  {journal} {Reports on Progress in Physics}\ }\textbf
  {\bibinfo {volume} {74}},\ \bibinfo {pages} {124502} (\bibinfo {year}
  {2011})}\BibitemShut {NoStop}%
\end{thebibliography}%

\end{document}